# About the Phasor Pathways in Analogical Amplitude Modulations

H.M. de Oliveira[1], F.D. Nunes[1]

**ABSTRACT**

*Phasor diagrams have long been used in Physics and Engineering. In telecommunications, this is particularly useful to clarify how the modulations work. This paper addresses rotating phasor pathways derived from different standard Amplitude Modulation Systems (e.g. A3E, H3E, J3E, C3F). A cornucopia of algebraic curves is then derived assuming a single tone or a double tone modulation signal. The ratio of the frequency of the tone modulator ($f_m$) and carrier frequency ($f_c$) is considered in two distinct cases, namely: $f_m/f_c<1$ and $f_m/f_c \geq 1$. The geometric figures are some sort of Lissajours figures. Different shapes appear looking like epicycloids (including cardioids), rhodonea curves, Lemniscates, folium of Descartes or Lamé curves. The role played by the modulation index is elucidated in each case.*

Keyterms: *Phasor diagram, AM, VSB, Circular harmonics, algebraic curves, geometric figures.*

## 1. Introduction

Classical analogical modulations are a nearly exhausted subject, a century after its introduction. The first approach to the subject typically relies on the case of transmission of a single tone (for the sake of simplicity) and establishes the corresponding phasor diagram. This furnishes a straightforward interpretation, which is quite valuable, especially with regard to illustrating the effects of the modulation index, overmodulation effects, and afterward, distinctions between AM and NBFM. Often this presentation is rather naïf. Nice applets do exist to help the understanding of the phasor diagram [1- 4]. However ...
Without further target, unpretentiously, we start a deeper investigating the dynamic phasor diagram with the aim of building applets or animations that illustrates the temporal variation of the magnitude of the amplitude modulated phasor. We decided to investigate the pathway described in the Argand-Gauss plane by the tip of the rotating phasor. Surprisingly, in contrast to that usually offered, it was found the outline of cornucopia (a galaxy of geometric figures associated with algebraic curves [5]) of geometric layouts, a polymorphism carrying a considerable visual richness. These findings can also be useful for people of design, technical drawing, illustrations etc... Lissajous figures are patterns generated by the junction of a pair of sinusoidal waves with axes that are perpendicular to one another. So are also these phasor diagrams, which are generate from in quadrature AM waves. In 1982, the ITU (International Telecommunication Union) designated the types of amplitude modulation [6], among them: A3E (double-sideband a full-carrier), H3E (single-sideband full-carrier), J3E (single-sideband suppressed-carrier), and C3F (vestigial-sideband). In engineering and applied science, sinusoid signals are the basis for studying most systems. The sinusoid is an idealized signal that models excitations that oscillate with a regular frequency (like AC power, AM radio, pure musical tones, and harmonic vibrations). Phasors are a natural tool for these frameworks.

---

[1] Federal University of Pernambuco, Centro de Tecnologia e Geociências, Brazil, HMdO,qPGOM, e-mail: {hmo,fdnunes@ufpe.br}, url: http://www2.ee.ufpe.br/codec/deOliveira.html

## 2. On phasor trajectories in the strict sense AM

A. case of dynamic phasor of a modulated single tone

The simplest analog modulation is AM. Trajectory of the tip of phasor resulting from a harmonically amplitude modulated oscillation (AM) for single tone [7]: control parameters are $\{m, f_m/f_c\}$, respectively, the index modulation AM and frequency of the tone relative to the carrier frequency. The (normalized carrier frequency) phasor equation $(x(t),y(t))$ for a single tone is given by

$$x(t)= \Re\{Z(m,f_m,t)\} \text{ and } y(t)= \Im\{Z(m,f_m,t)\}, \qquad (1)$$

where

$$Z(m, fm, t) := A \cdot \left(1 + \frac{m}{2} \cdot e^{\sqrt{-1}\cdot 2\cdot \pi \cdot fm \cdot t} + \frac{m}{2} \cdot e^{-\sqrt{-1}\cdot 2\cdot \pi \cdot fm \cdot t}\right) \cdot e^{\sqrt{-1}\cdot 2\cdot \pi \cdot t}. \qquad (2)$$

Some diagrams for a single tone A3E modulation are shown in Figs.1-4 by increasing the modulation index. These diagrams have also likeness to the eye diagrams of digital communications [8]. Here, the modulation index $m$ plays a role to close the eye as it can be seen from Fig.1 to Fig.4 (rising $m$ values). A gif animation is available at the URL http://www2.ee.ufpe.br/codec/Transition0v5.gif for a pre-set single tone.

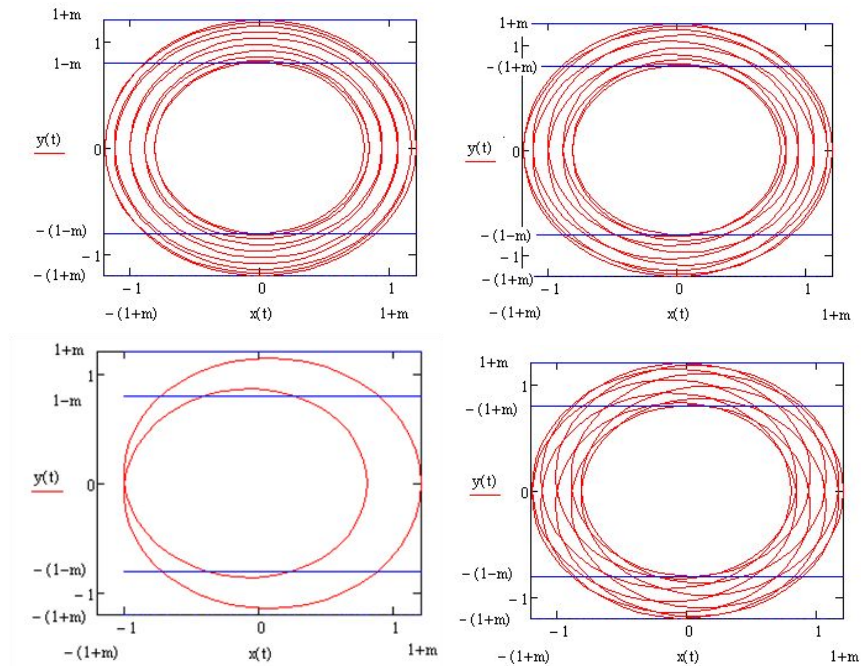

Fig.1 A3E-Modulation Parameters- $m$=0.2 and frequency ratio $f_m/f_c$: a) 0.1 b) 0.3 c) 0.5 d) 0.7.

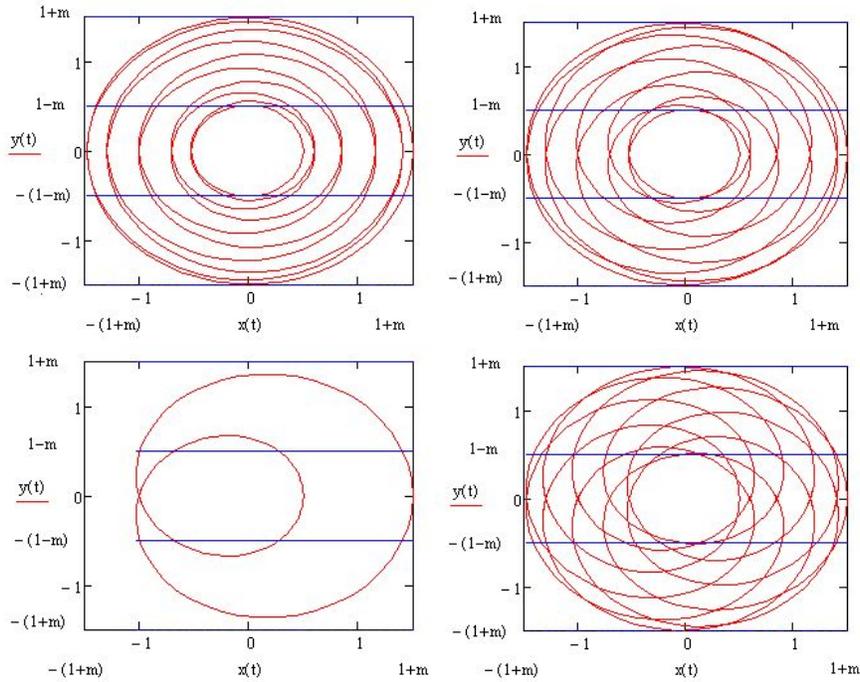

Fig.2 A3E-Modulation Parameters- $m$=0.5 and frequency ratio $f_m/f_c$: a) 0.1 b) 0.3 c) 0.5 d) 0.7.

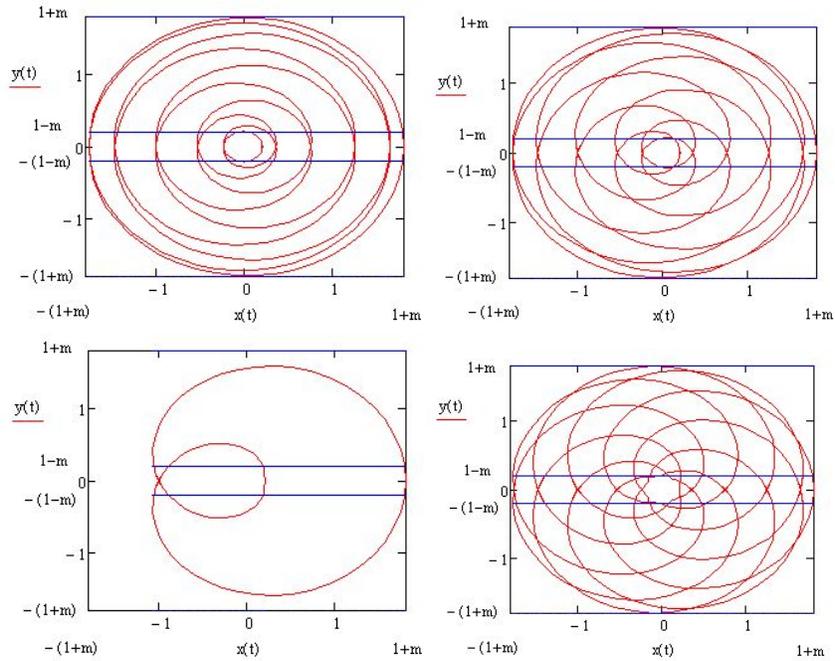

Fig.3 A3E-Modulation Parameters- $m$=0.8 and the frequency ratio $f_m/f_c$: a) 0.1 b) 0.3 b) 0.5 d) 0.7.

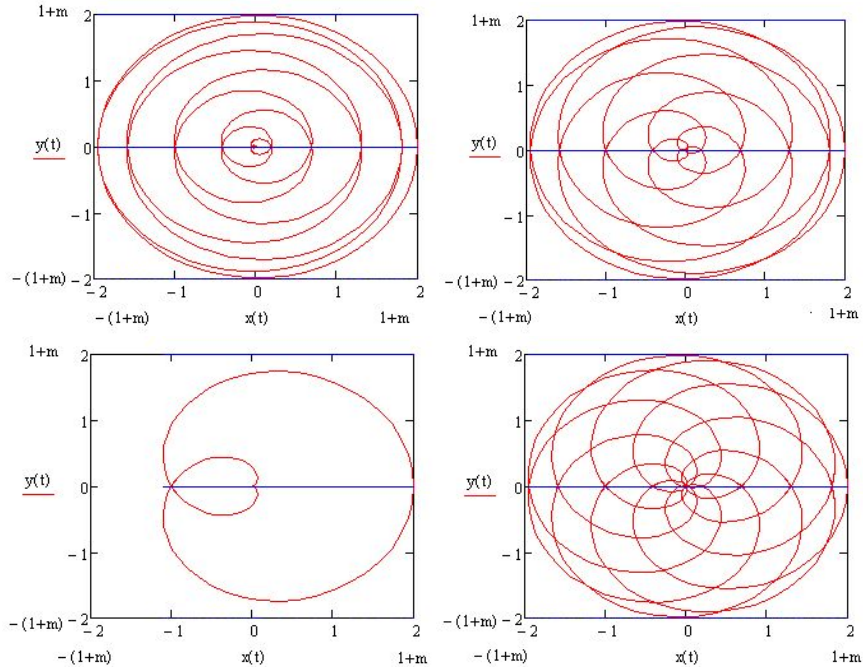

Fig.4 A3E-Modulation Parameters- (closed eye) *m*=1,0 and the frequency ratio $f_m/f_c$: a) 0.1 b) 0.3 c) 0.5 d) 0.7.

Lissajours figures are patterns generated by the junction of a pair of sinusoidal waves with axes that are perpendicular to one another. So are also these phasor diagrams, which are generate from in quadrature AM waves. In effect, in order to display such phasor diagrams in a screen of an oscilloscope, an arrangement quite similar to the Hartley phase shift method for SSB modulation [**9**] can be used, as illustrated in Fig.5.

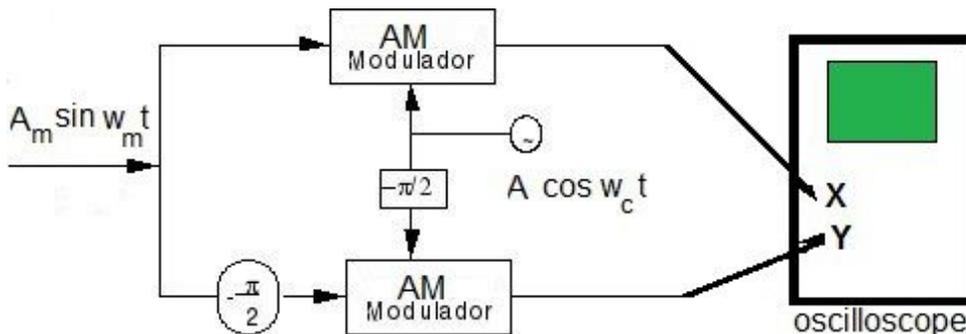

Fig.5 Phasor pathways generation with oscilloscope in X-Y mode. Both (identical) AM modulators can be A3E, H3E, J3E... A Hilbert filter is required.

Cases in which the frequency of the modulating signal $f_m$ is greater than the carrier $f_c$ generates pathways of particular concern (even if the ratio $f_m/f_c$>1is not of usual interest in telecom). If the relationship is exact and the modulation index is unity, one has "rosettes" figures (Fig.7) with exactly the same number of lobes to this ratio (rose or *rhodonea curve*). These curves are some sort of different *epicycloids* [**10-11**]. When the modulation tone has a frequency that is twice the frequency of the carrier, curves of the phasor tip bear a resemblance to a *Lemniscate of Bernoulli* and *Nephroid*s with the modulation depth (AM modulation index) playing a role to "tighten/slacken" the "knot". A gif animation to illustrate this effect is available at the URL http://www2.ee.ufpe.br/codec/Transition2.gif.

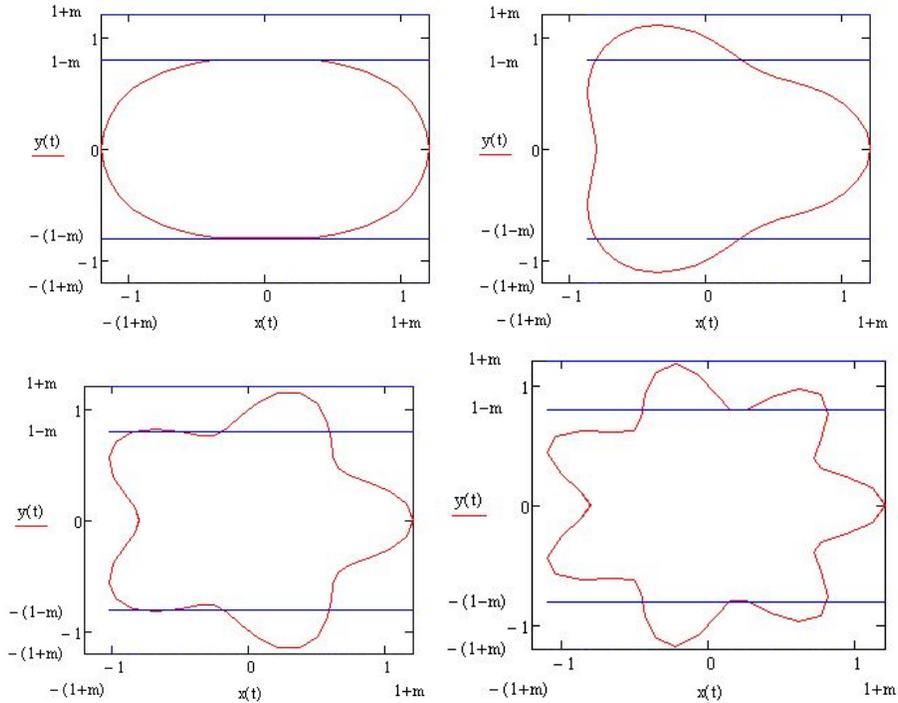

Fig.6 Parameters: $m=0.2$ $f_m/f_c=$ a) 2 b) 3 c) 5 d) 7.

Simple gif animations for clarifying the role of the modulation index are available at the URL http://www2.ee.ufpe.br/codec/Transition3.gif and http://www2.ee.ufpe.br/codec/Transition5.gif
The "*petals*" of the curves are spaced at an angle of $2\pi/M$, where one is at 0 rad.

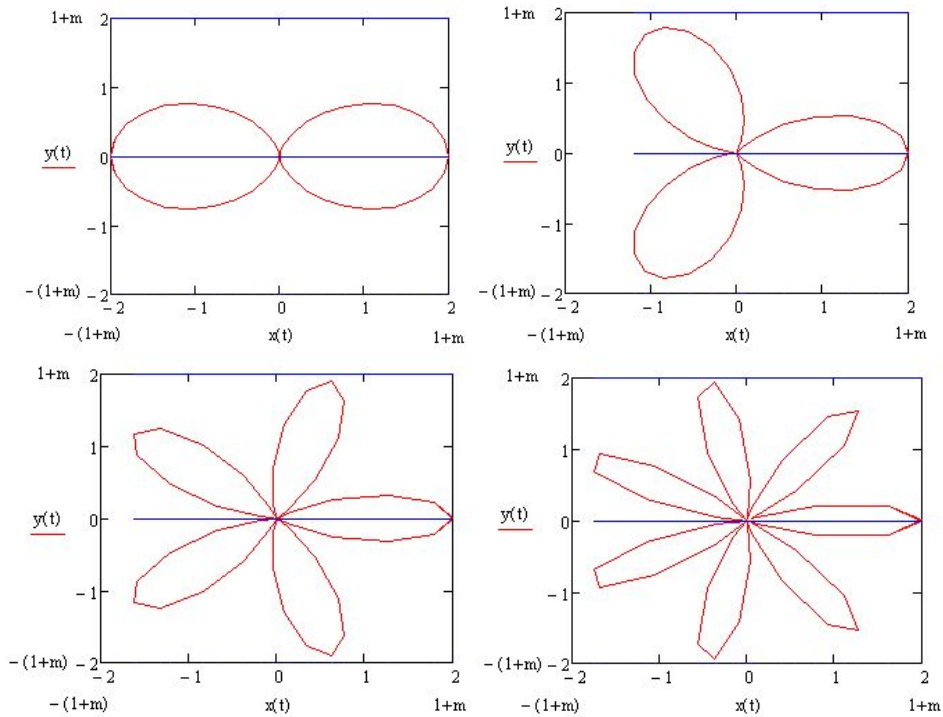

Fig.7 Rhodonea-like curves. Parameters: $m=1$ and the frequency ratio $f_m/f_c$: a) 2 b) 3 c) 5 d) 7.

B. Pathways for musical notes into an octave

In order to visually illustrate the relationships between musical notes in the diatonic scale (Ancient Greek), consider the fixed carrier frequency in a C of a particular octave (denoted by $f_{do}$), and another individual note (Re, Mi, Fa, Sol, La, Si), in the same octave applied in wave amplitude modulated. The signal whose phasor diagram is shown corresponds to an AM waveform $[1+\cos(2\pi f_{note} t)] \cdot \cos(2\pi f_{do} t)$.

| *musical note* | Do | Re | Mi | Fa | Sol | La | Si | |
|---|---|---|---|---|---|---|---|---|
| | 264 | 297 | 330 | 352 | 396 | 440 | 495 | (Hz). |

The figures are similar for all octaves. The idea is to compare with the Von Helmholtz curve of consonance-dissonance [**12**]. Note that the unison is related to a *cardioid curve* (*epicycloid*).

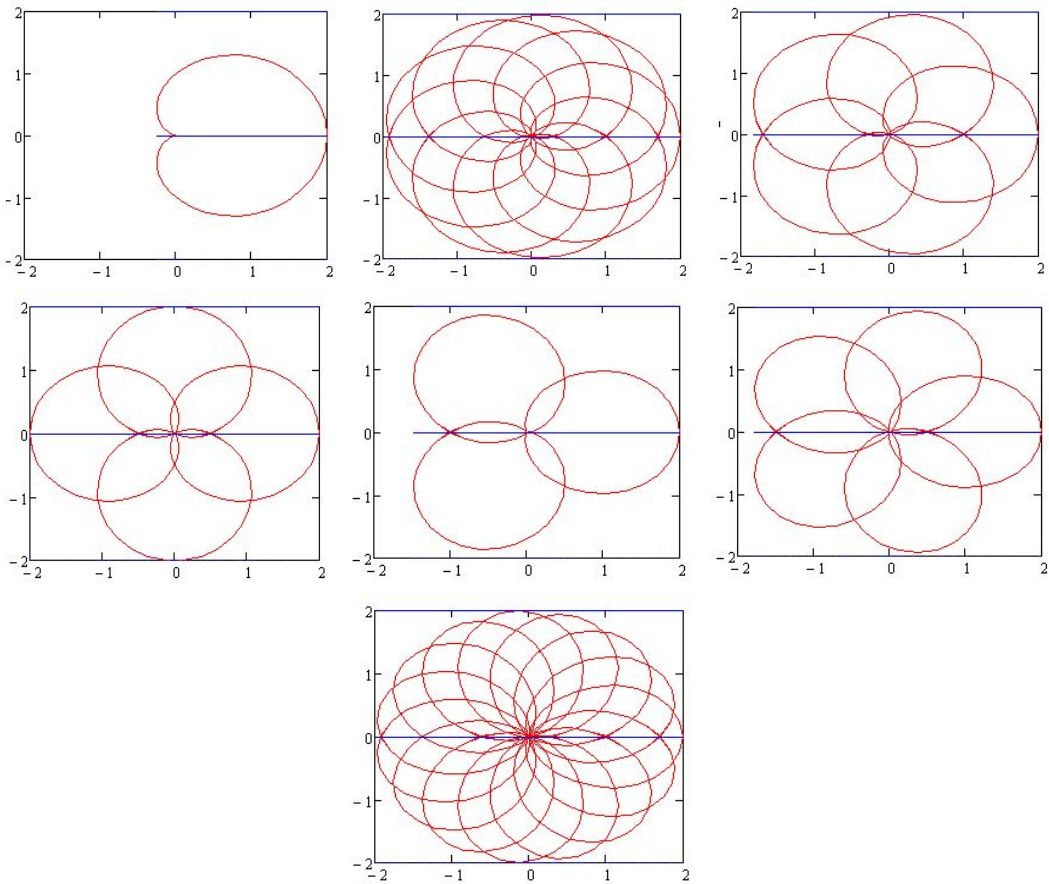

Fig.8 Phasor pathways for a fixed musical note (*do*) and further notes of the same octave.

C. case of dynamic phasor of modulated bi-tones

The tip of the phasor is given by a two-parameter Eq(2), but not Eq(1) is replaced by:

$$Z(m1, m2, fm1, fm2, t) := \left[ 1 + \left( \frac{m1}{2} e^{\sqrt{-1} \cdot 2\pi \cdot fm1 \cdot t} + \frac{m1}{2} e^{-\sqrt{-1} \cdot 2\pi \cdot fm1 \cdot t} \right) + \left( \frac{m2}{2} e^{\sqrt{-1} \cdot 2\pi \cdot fm2 \cdot t} + \frac{m2}{2} e^{-\sqrt{-1} \cdot 2\pi \cdot fm2 \cdot t} \right) \right] \cdot e^{\sqrt{-1} \cdot 2\pi \cdot t} \quad (3)$$

A few examples of geometric figures built by bi-tones phasor pathways are presented in the sequel, for instance, assuming two sinusoidal tones with $f_{m1}/f_c=5$ and $f_{m2}/f_c=0.1$. The modulation indexes are then varied (Fig.9).

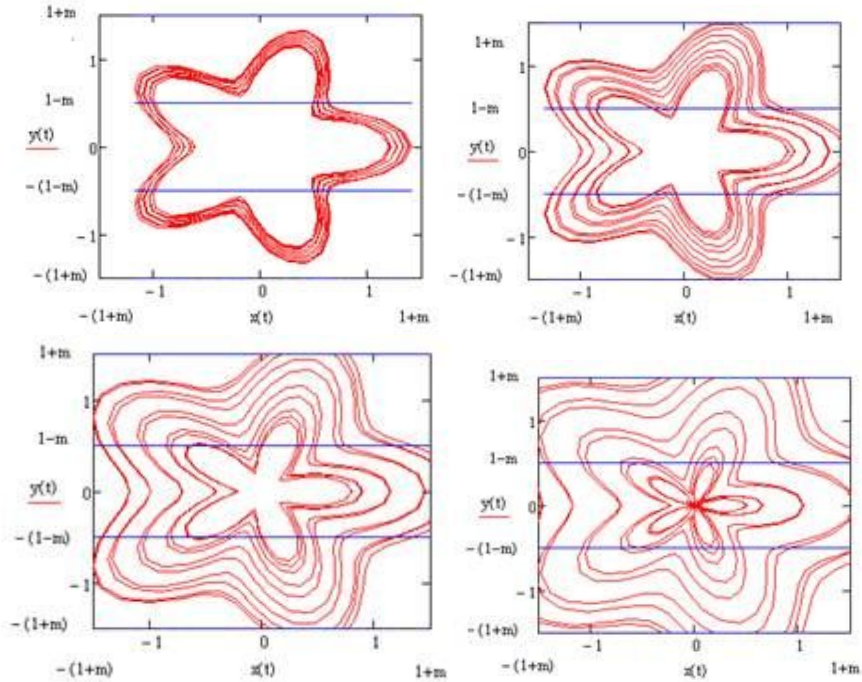

Fig.9 Phasor diagram for a couple of tones, low carrier frequency.
a) $m_1=0.3$ $m_2=0.1$ b) $m_1=0.3$ $m_2=0.3$ c) $m_1=0.3$ $m_2=0.5$ d) $m_1=0.3$ $m_2=0.8$.

Let us now consider another bi-tone example (for instance, by assuming always $m_1=0.5$ and $m_2=0.2$), but now varying the ratio between the frequencies of the two sinusoidal tones (Fig.10).

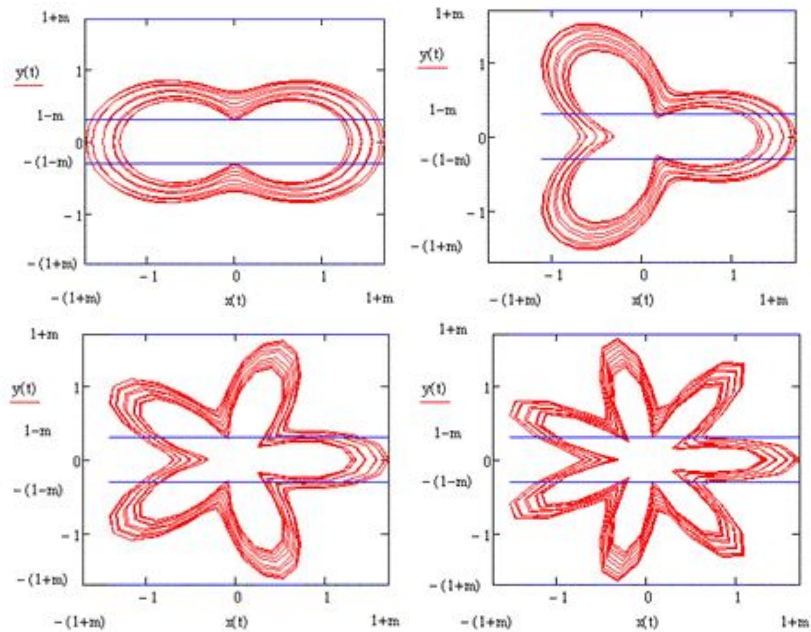

Fig.10 Phasor trajectory of two tones of frequencies $f_{m1}$ & $f_{m2}$ AM-modulated. The chosen (normalized frequencies) are: a) 2 & 0.1 b) 3 & 0.1 c) 5 & 0.1 d) 7 & 0.1.

## 3. Phasor trajectories in the broad-sense AM

A. The case of dynamic phasor of single sideband modulation

To face both double and single side band AM, we plot the phasor pathways of AM and SSB (carrier suppressed or not). The expected was dealing with a slow frequency tone modulating a high frequency carrier. For a single tone, we consider now the complex given by

$$Z(\alpha, m, fm, t) := \left[ A + \frac{\alpha}{2} \cdot m \cdot e^{\sqrt{-1} \cdot 2 \cdot \pi \cdot fm \cdot t} + \left(1 - \frac{\alpha}{2}\right) \cdot m \cdot e^{-\sqrt{-1} \cdot 2 \cdot \pi \cdot fm \cdot t} \right] \cdot e^{\sqrt{-1} \cdot 2 \cdot \pi \cdot t} \quad (4)$$

where the parameters $\alpha \in \{0,1\}$ corresponds to single and double side band, respectively, and the value of the amplitude $A \in \{0,1\}$ accounts for suppressed carrier or not, respectively. Surprisingly, the "vertices" of many SSB-pathway are often some kind [11] of *folium of Descartes, Tschirnhaus's Cubic* or *Conchoid of de Sluze* (Fig.11-13). However, the exact relationships with known curves are not on the scope of this primary approach.

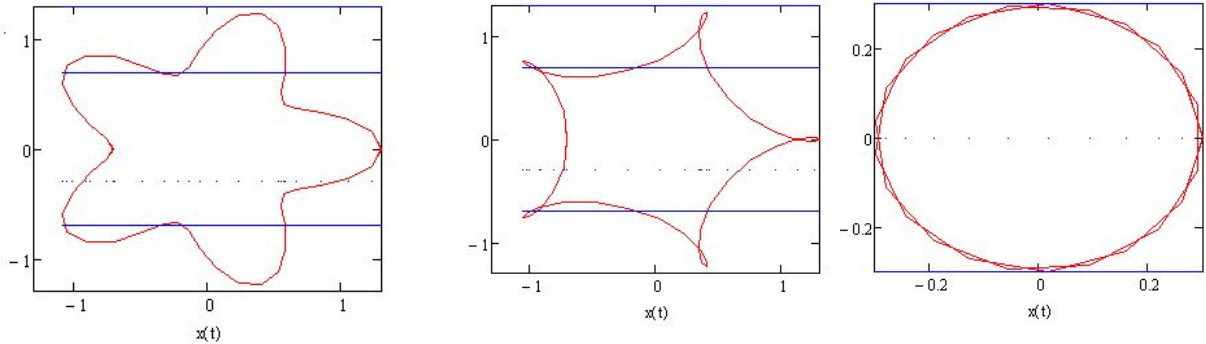

Fig.11 $f_m/f_c$=5, $m$=0.3 a) A3E-modulation AM $\alpha$=1, b) H3E-Modulation SSB+C $\alpha$=0, c) J3E-Modulation SSB $\alpha$=0.

To gain some insight into the effects of the modulation index in a single tome modulation, we consider only SSB+C (the effects are much apparent), and gradually increase the value of *m*.

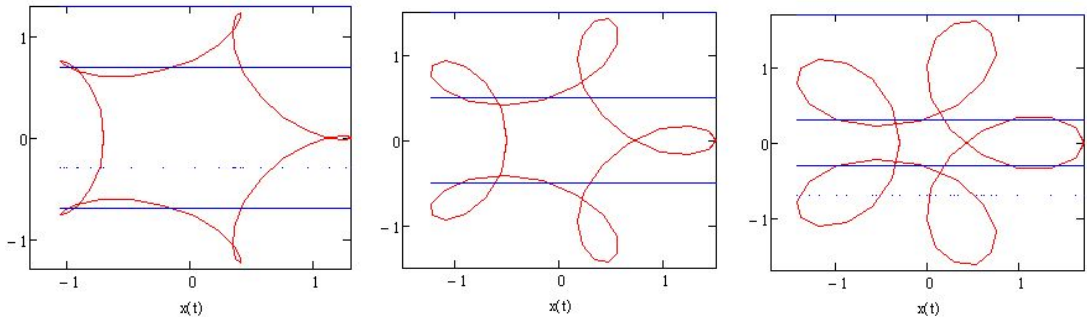

Fig.12 H3E-Modulation with $f_m/f_c$=5, SSB+C a) $m$=0.3, b) $m$=0.5, c) $m$=0.7.

B. The case of dynamic phasor of vestigial sideband modulation

In order to glimpse the phasor envelope for Vestigial Side Band modulation (C3F), we consider a simple case of a single tone, setting the modulation index $m$=0.3 and assuming a slow carrier (with is indeed unusual in practical telecommunication applications). This can be achieved by considering a real value for $\alpha$, constrained in the range $0 \leq \alpha \leq 1$. We start with a low residue (compare with SSB, Fig.12a) and gradually increase the factor $\alpha$ towards AM, i.e. unity (AM double side band, Fig.11a). As in earlier cases, some resemblance can be found to further interesting known algebraic curves such as *Lamé's superellipse* ($n=1/2$, $a=b=1$).

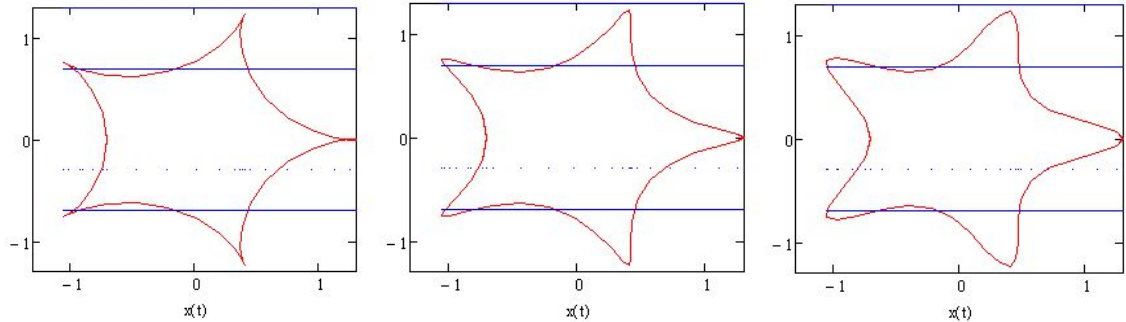

Fig.13 C3F-Modulation (VSB) with $f_m/f_c=5$, $m=0.3$ and different degrees of vestigial side bands:
a) $\alpha=0.1$, b) $\alpha=0.3$, c) $\alpha=0.5$.

### C. Drift Analysis and Irrational Envelope-carrier Frequency Ratios

Small perturbations can occasionally happen in the ratio $f_m/f_c$ due to frequency drifts. In order to investigate their effects, a number of phasor trajectories are plotted as illustration. It is considered an integer ratio $f_m/f_c=M$ with a random perturbation confined within 10%.

$$\tag{5}$$

where $\delta$ is a uniform random variable with distribution $\delta \sim U(0,1)$. Fig.14 shows the pathways for $M=3$ and modulation index $m=0.2$.

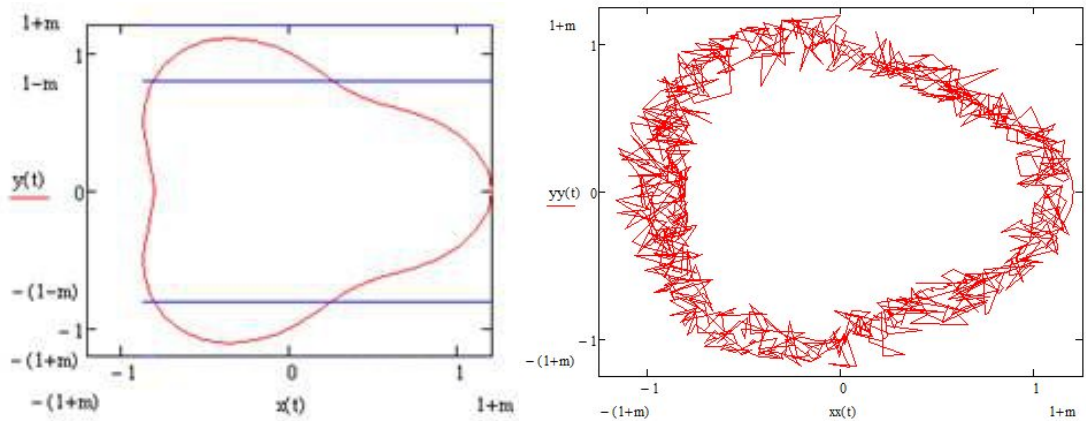

Fig.14 (a) Phasor pathways for a single tone $f_m=3f_c$ ($M=3$) and modulation index $m=0.2$,
(b) idem, now with a 10% random frequency drift (random fluctuation).

Another investigation deals with envelope-to-carrier frequency ratio. As expected, non periodic pathways are generated when this ratio is irrational. Particularly, the case $f_m/f_c=$ is plotted.

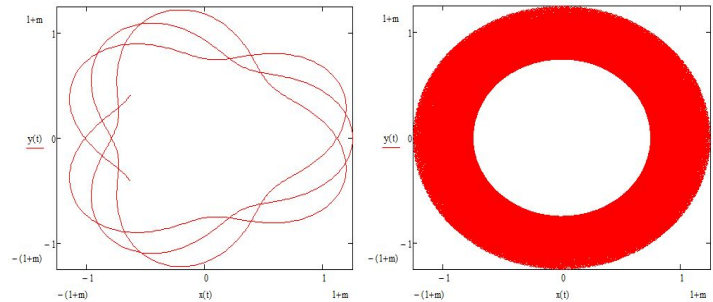

Fig.15 Phasor pathways for irrational envelope-to-carrier frequency ratio: $m$=0.25, $f_m/f_c$= (a) initial pathways that never close; (b) long-run pathway tends to fulfill the ring of radius from 1-$m$ to 1+$m$.

## 4. Conclusions

It seems a bit odd that there is so much unexplored wealth in a subject so classic and well established as the phasor diagrams of amplitude modulations. One potential application for such geometric patterns is to provide electronic lighting walls or scenarios during live shows of bands or musical performances. Patterns can be driven by the piece of music performed, providing thus rich visual effects. Although this presentation does not contain any original scientific, the approach is welcoming, valuable for understanding the modulations and can provide extra motivation for engineering students and technicians. The central goal is therefore within the scope of the transactions on education. These findings can as well be useful for people of the fields of design and technical drawing etc. For completeness, it is suggested the mainstreaming of this tool when presenting the AM systems.